
\magnification 1200
\tolerance=10000
\hsize=15.5 true cm
\hoffset=1 true cm
\baselineskip=2\baselineskip

\centerline{\bf SELF-CONSISTENCE OF SEMI-CLASSICAL GRAVITY}
\centerline{Wai-Mo Suen}
\centerline{McDonnell Center for the Space Sciences}
\centerline{Department of Physics, 1105}
\centerline{Washington University, One Brookings Drive}
\centerline{St. Louis, MO 63130-4899}
\bigskip
\centerline{\bf Abstract}

Simon argued that the semi-classical theory of gravity,
unless with some of its solutions excluded,
is unacceptable for reasons of both
self-consistency and experiment, and that it has to be replaced
by a constrained semi-classical theory.  We examined whether the
evidence is conclusive.
\vskip 3 true cm
\noindent
\vfill\eject

Recently Simon$^1$ suggested that the semi-classical theory of
gravity in which the spacetime metric is treated as a
classical object coupled to quantum fields,$^2$
is self-inconsistent unless some of its solutions are excluded.
The point is that the
semi-classical Einstein equation admits solutions which are not
perturbatively expandable about classical solutions (solutions
satisfying the classical Einstein equation).  Simon called these
nonperturbatively expandable solutions pseudosolutions.  He
claimed that the semi-classical theory is self-{\bf in}consistent
unless the pseudosolutions are excluded by the introduction of
perturbative constraints.$^1$

The reasoning behind the claim is the following:  The
semi-classical Einstein equation is
$$ G_{\mu\nu} = 8\pi G < \hat T_{\mu\nu} >^{\rm ren}\quad , \eqno (1)$$
where $G_{\mu\nu}$ is the Einstein tensor and
$< \hat T_{\mu\nu} >^{\rm ren}$ is
the renormalized expectation value of the energy-momentum tensor
of the quantum fields.  It is noted that$^1$ the calculation of
$< \hat T_{\mu\nu} >^{\rm ren}$ involved an expansion in power of $\hbar$
$$
    < \hat T_{\mu\nu} >^{\rm ren}
  = < \hat T_{\mu\nu} >_0^{\rm ren}
  + < \hat T_{\mu\nu} >_1^{\rm ren} \hbar
  + O( \hbar^2 ) \quad . \eqno (2)
$$

The one-loop terms $< \hat T_{\mu\nu} >_1^{\rm ren}$, contains higher
derivatives, $\sqcup R$, $\nabla_\mu\nabla_\nu R$, etc.
They have time derivatives of order higher than those of the
$G_{\mu\nu}$ on the left hand side of (1).  By a dimensional
argument, the $O(\hbar^2$) term involves even higher
derivatives.  Now if one insert (2) into (1) with the
$O(\hbar^2)$ terms truncated, the higher derivative in the
one-loop term will drastically change the character of Eq.~(1),
independent of the smallness of $\hbar$.  The solution space is
much larger than that of the classical Einstein equation
$(\hbar =0)$, in particular, there are solutions for which the higher
derivative contributions are as large as the classical
contributions, and hence not perturbatively expandable about the
classical solutions.  They are the ``pseudosolutions'' of
Ref.~1.  Apparently for these solutions the expansion (2) breaks
down and the truncated expression should not be used as the source
term in (1) in the first place, as presumably the even higher
derivative terms $O(\hbar^2)$ in (2), which has been dropped,
have significant contribution.  Hence the semi-classical theory
in its standard formulation, {\it i.e.,} a formulation with
the ``pseudosolutions'' included, seems to be self-inconsistent.

Reference 1 further noted that flat space is unstable in the
semi-classical theory due to the existence of the higher
derivative terms,$^{3-6}$  and argued that this means,
independent of the derivation, semi-classical theory
in its standard formulation is unacceptable,
even if one is to postulate it.
It was concluded that:$^1$  ``So, for two reasons,
self-consistency and experiment, we should consider perturbative
(constrained) semiclassical theory as the ``correct''
semiclassical gravity, or at least as a potentially correct
theory.  Semi-classical gravity that does not exclude
pseudosolutions cannot be considered even a potentially correct
theory.''

In the following we examine whether the evidence is conclusive.
We first look at the case of free fields for which the
semi-classical theory is very well-studied.$^2$  The energy-momentum
tensor operator is quadratic in the field operators$^7$
$$\hat T_{\mu\nu} = \hat \phi_{;\nu} \hat \phi_{;\nu}+\ldots \eqno (3)$$
The expectation value $<\hat T_{\mu\nu}>$ given by (3) with
respect to any quantum state is divergent.  Nevertheless we can
calculate it explicitly, by, say, point-splitting.$^2$  Notice
that the resulting quantity (i)~contains higher derivative terms
and (ii)~is not an expansion in $\hbar$.

The next step is to obtain a finite object from this divergent
quantity.  We have to determine what divergent part to subtract
off.  It is in figuring out what to subtract that we use an
expansion.  The counter terms $< T_{\mu\nu} >^{\rm DS}$ is taken
as the first few terms of this expansion, and the renormalized
energy-momentum tensor is given by
$$ <T_{\mu\nu}>^{\rm ren} = <T_{\mu\nu}> -
   <T_{\mu\nu}>^{\rm DS}\quad . \eqno (4)$$
This is the essence of many renormalization schemes$^2$ based on
the DeWitt-Schwinger expansion.$^8$  In particular, in adiabatic
regularization,$^9$ $<T_{\mu\nu}>^{\rm ren}$ is explicitly
represented in this form.  For example, in the case of massless, conformally
coupled scalar field in Robertson-Walker spacetime studied in
Ref.~1, the right hand side of (4) in terms of the adiabatic
regularization, is$^{10}$
$$ \eqalign{ <T_{tt}> =
   & {1 \over 2880\pi^2} \left [ 3H^4+H\dot R -
     {1\over 12} R^2+H^2R \right ] \cr
   &+ {1 \over 4\pi^2} \int_0^\infty {k^2dk \over a^3}
     \left \{ a \vert \dot \psi_k \vert^2
     + {k^2 \over a} \vert \psi_k \vert^2 \right \} \cr}\eqno (5)$$
$$ <T_{tt}>^{\rm DS} = {1 \over 4\pi^2}
   \int_0^\infty {k^2dk \over a^3}
   \left ( {k \over a} \right ) \quad . \eqno (6)$$
\noindent
$\{$We display only the $tt$ component for the spatially flat metric;
$a$ is the scale factor, $H=\dot a /a$ is the Hubble parameter,
$R$ is the scalar curvature, an over dot is differentiation with
respect to the proper time.  $\psi_k(t)$ is the mode function of
the scalar field
$\hat\phi = \int (d^3k/(2\pi)^{3/2})[\hat a_k(e^{ik\cdot x}/a)\psi_k(t)
  +\hat a_a^+ (e^{-ik\cdot x}/a)\psi_k^*(t)]\}$.

Two comments of this procedure of calculating $< T_{\mu\nu}>^{\rm ren}$
are in order.  Firstly, indeed an expansion is involved in
the calculation, but only in the determination of the counter
terms $< T_{\mu\nu}>^{\rm DS}$.  It is {\bf not} that if we want
higher accuracy, we have to include more terms in
$< T_{\mu\nu}>^{\rm DS}$.  So it is {\bf questionable} whether we
have to view the resulting $<T_{\mu\nu}>^{\rm ren}$ as just the
leading terms of an expansion, and that higher and higher
derivative terms have been neglected.

What if one choose to include more terms in the counter piece
$<T_{\mu\nu}>^{\rm DS}$?  One is free to include higher and
higher order derivative terms as long as they are finite, but
there is {\bf no} logical reason why one has to do that.  The
``traditional'' choice of the counter piece can be viewed as the
``minimal subtraction'', as called by Parker.$^{11}$  In this
point of view, the renormalized energy-momentum tensor in
semi-classical gravity with minimal subtraction is {\bf not} the first
few terms in an expansion containing higher and higher derivative
terms.

The second comment is more philosophical than logical.
Independent of what one choose for the counter piece, whether
minimal subtraction or not, the ``bare'' piece $<T_{\mu\nu}>$
[first term on right hand side of (4)] by itself contains higher
derivative terms, which is not coming from any expansion.  Nature
could be telling us something by this and we should not easily
dismiss this piece of information.

The above discussions are for free fields.  For interacting
fields, indeed in general we can only calculate the ``bare''
piece $<T_{\mu\nu}>$ by expansion, but then we can imagine with the
developments of the non-perturbative field theory, we might
someday be able to calculate $<T_{\mu\nu}>$ without resorting to
any expansion (in the self-coupling parameter or in
$\hbar$).  The resulting non-perturbative $<T_{\mu\nu}>$ must also
contain higher derivative terms, as it has to reduce to the free
field case in vanishing coupling.  In this sense the above
discussions might not be crucially relying upon the field being a
free field.

Obviously, without a full quantum theory of gravity, there is no
way to tell for sure if the standard semi-classical gravity is a
reasonable approximate description of nature in some region of
superspace.  What we want to argue here is that,
within the context of semiclassical gravity ({\it i.e.,}
classical spacetime coupled to quantum fields), it is {\bf not}
necessarily logically inconsistent to consider the full effect of
the higher derivative terms in the theory.
We want to further mention that in many other cases, {\it e.g.,}
in the post-newtonian expansion of the equation of motion for
particles in general relativity, higher derivative terms arise as
a result of an expansion.$^{12}$  In such cases, it is truly
logically inconsistent to consider the ``pseudosolutions'', and
the ``contrained treatment'' as advocated by Simon$^1$ should be
used.  In Ref.~12, such a constrained treatment was carried
out for the post-Newtonian expansion.
However, in semi-classical gravity, logical consistency
does not immediately imply such a treatment.  This leads us to study
the next question:  Is the higher derivative terms
with its ``pseudosolutions'' compatible with experiment?

Reference 1 suggested that the instability of flat space as shown
in Refs.~3--6 implied incompatibility.  However, for the
instability studied in Refs.~3 and 4, the unstable time scale
involves an unknown parameter which can be made arbitrarily long
and hence the instability may not be practically noticeable, as
pointed out in Ref.~3.  On the other hand, if the time scale is
too short ({\it e.g.,} of order Planck), we do not expect the
semi-classical treatment to be valid in the first place, and
hence such an instability becomes irrelevant to the consideration of
whether the ``pseudosolutions'' are physical.$^{13}$
Further more, the stability study of the
Minkowski space in Refs.~3 and 4 are carried out only to the
linear order.  As also pointed out in Ref.~3, an unstable solution
to the linearized equation might not really indicate instability of
the full equation.  We note that this is particularly the case in
the study of instability about the Minkowski space, as the
initial constraint equation (the $t-t$ component of the
semi-classical Einstein equation) when expanded about the
Minkowski solution is second order in the perturbation, and is
hence identically zero to the linear order.  The existence of an
unstable linearized solution may have no physical significance if
its second-order extension does not satisfy the initial
constraint equation.  In Refs.~5 and 6, unstable modes, which are
isotropic in space, are found to exist for the full semi-classical
Einstein equation independent of the sign and value of the
renormalization parameters; and the unstable time scales can be
arbitrarily picked.  So the
Minkowski space is indeed catastrophically unstable for such
modes.  But as pointed out in Ref.~6, for spatially isotropic
perturbation, this instability of the Minkowski space might not be
significant until a time in the far future:  For isotropic
perturbation, the relevant background should be the Friedman
universe instead of the Minkowski space.  For the particular
unstable modes considered, the Hubble parameter of the present
universe is still large.  It  is further shown in Ref.~6 that the
Friedmann universe is stable under infinitesimal perturbations in
the semi-classical theory.  Hence at present we do not have solid
physical evidence on experimental/observational ground to rule
out the standard semi-classical gravity.

One might argue that the fact that all observed phenomena are
describable by the classical Einstein equation is an evidence
against the standard semi-classical theory:  With the
higher derivative terms in the theory, the solution space is much
larger, with a large class of solutions which are not
approximately classical solutions (pseudosolutions of Ref.~1).
Why do we not see such solutions in the universe?  We would
like to suggest the possibility that the semi-classical theory,
for a range of initial condition, with a set of realistic
quantum fields included, may have a set of classical solutions
(solutions to the classical Einstein equation) as attractors.  A
heuristic argument for that is:  When the higher derivative terms
dominate the semi-classical equation, there are usually rapid
changes in the spacetime geometry.  Such rapid change would excite
the quantum fields and the back-reactions can make the higher
derivative terms become subdominant, at least for some regime in
the parameter space.  An explicit example of this is the
reheating of the higher derivative inflationary models (which
includes the Starobinsky model$^{14}$ criticized in Ref.~1).  It is
shown$^{15}$ that for a wide range of initial parameters the
universe ends up in an oscillation phase (a ``pseudosolution''),
the changing geometry excites the conformally {\bf non-incovariant}
quantum fields in the universe, and the evolution as determined by
the semi-classical equation automatically drives the spacetime
into a radiation dominated Friedmann solution$^{16}$ ({\it i.e.,} a
``pseudosolution'' automatically turns into a classical
solution).  Such an automatic transition from dominated by the
higher derivative terms to dominated by the classical Einstein
terms is not possible if only conformally invariant quantum fields
are included in the model, as is in the case studied in Ref.~1.
Indeed as shown in Ref.~6, both radiation dominated and
matter-dominated Friedmann universes are unstable to
infinitesimal perturbation in some measure, if only conformally
invariant fields are considered.  However, in comparing the
semi-classical theory to actual experiment/observation, we should
use a realistic set of quantum fields.  We believe,
whether the standard semi-classical theory,
in a certain range of renormalization parameters, with a
realistic set of fields, is consistent with experiment/observation
or not is still an open
question.

There is one related issue that we want to comment on.  In the
discussion above, we argue that, within the semi-classical theory,
which treats spacetime as a classical object coupled to quantum
fields, it might not be self-inconsistent to consider the effects
of the higher derivative terms in full.  One can go further and
ask:  We know that classical spacetime is most likely the
$\hbar \rightarrow 0$ limit of some quantum spacetime
description; hence (i)~Is it meaningful to retain the
$\hbar \ne 0$ effects coming from the matter fields?  (ii)~If yes, would
this restrict the significance of the higher derivative terms?

The first question has been discussed extensively in the
literature.$^2$  Without a full quantum theory of gravity, there
is clearly no way to draws a definite conclusion.  What we want
to note here is that it is misleading to simply count the power of
$\hbar$ in demanding consistency.  It could be consistent to
study different subsystems to different orders of $\hbar$, even
when these subsystems are coupled.  An elementary example is the
Born-Oppenheimer approximation in molecular physics, where the
nuclei are treated to zeroth order in $\hbar$, while electrons
are to all orders.  In the case of the yet unknown full quantum
theory of spacetime and matter fields, the Planck mass could have
the role of nucleus mass, and the semi-classical theory could be
meaningful in some region of the superspace.$^{17}$  This
implies that some of the solutions
dominated by the higher-derivative terms in the semi-classical
theory (``pseudosolution'')
are not physically meaningful in the context of the full
theory, namely, those solutions involving Planck energy and
time scales.  However, not all ``pseudosolutions'' involve the
Planck scale.  For example, the unstable modes studied in
Refs.~5, 6, as infinitesimal perturbations of the
Minkowski space is as far away from the Planck scale as anything
can be, at least for a period of time whose length depends on
the choices of the initial parameters.  In the traditional wisdom
that quantum gravity comes in at the Planck scales, such
``pseudosolutions'' could have significance as approximations
in the context of a full quantum theory.

In conclusion, we argued that the semi-classical theory is not
necessarily self-inconsistent.  Further, some nonperturbatively expandable
effects of the higher derivative terms which do not involve the
Planck scale could be meaningful even in the context of a full
quantum theory.  We can even take the attitude that if the (yet
unknown) full quantum gravitational sector turns out to have no
higher derivative terms, the higher derivative terms generated by
the matter field sector could be very significant; whereas if
the full quantum sector does have higher derivative terms, our
present study of the higher derivative terms in the semi-classical
theory can be a useful warmup exercise.  It is also argued that
at present there is no solid observational/experimental evidence
against the inclusion of the ``pseudosolutions''.  Without
excluding such solutions, the semi-classical theory is surely much
more falsifiable, which is not necessary an undesirable feature.
It is hope that this paper can draw more attention on this issue
of observational/experimental evidence of semiclassical gravity.
\bigskip

I would like to thank Leonid Grishchuk, Leonard Parker,
Ian Redmont, and Jonathan Simon
for useful discussions.  This work was
supported in part by the National Science Foundation under Grant
No. 91-16682.
\vfill\eject
\centerline{\bf References}
\medskip
\item{1.}
J. Z. Simon, Phys. Rev. D {\bf 45}, 1953 (1992).
See also J.~Z.~Simon, Phys. Rev. D {\bf 41}, 3720 (1990);
{\bf 43}, 3308 (1991).
\item{2.}
See for example, N. D. Birrell and P.~C.~W. Davies,
{\it Quantum Fields in Curved Space} (Cambridge University Press,
Cambridge, 1982), and references therein.
\item{3.}
G. T. Horowitz and R. M. Wald, Phys. Rev. D {\bf 17}, 414 (1978);
G.~T.~Horowitz, {\it ibid} {\bf 21} 1445 (1980)
\item{4.}
S. Randjbar-Daemi, J. Phys. {\bf A14}, L229 (1981);
R.~D.~Jordan, Phys. Rev. D {\bf 36}, 3593 (1987).
\item{5.}
W.-M. Suen, Phys. Rev. Lett. {\bf 62}, 2217 (1989).
\item{6.}
W.-M. Suen, Phys. Rev. D {\bf 40}, 315 (1989).
\item{7.}
See, for example, Sec.~3.8 of Ref.~2, where the explicit
expressions of the energy-momentum tensors of free fields with
arbitrary spins are given.
\item{8.}
J. Schwinger, Phys. Rev. {\bf 82}, 664 (1951); B.~S.~DeWitt,
Phys. Rep. {\bf 19C}, 297 (1975).
\item{9.}
Adiabatic regularization was first introduced by L.~Parker and
S.~A.~Fulling, Phys. Rev D {\bf 7}, 2317 (1973).  It has been
shown to be equivalent to point-splitting by N.~D.~Birrell,
Proc. R. Soc. London {\bf B361}, 513 (1978), and P.~R.~Anderson
and L.~Parker, Phys. Rev. {\bf D36}, 2963 (1987) in the cases
studied.
\item{10.}
T. S. Bunch, J. Phys. {\bf A13}, 1297 (1980).
\item{11.}
L. Parker, private communication.
\vfill\eject
\item{12}
L. P. Grishchuk and S. M. Kopejkin, in {\it Relativity in
Celestial Mechanics and Astronomy}, ed. J.~Kovalevsky and
V.~A.~Brumberg (IAU. 1986).
\item{13}
Similar instabilities exist in QED and QCD.  For comparisons
between instabilities in semi-classical gravity and those in QED
and QCD, see R.~D.~Jordan, Phys.~Rev.~D {\bf 36}, 3593 (1987).
\item{14.}
A. A. Starobinsky, Phys. Lett. {\bf 91B}, 99 (1980).
\item{15.}
M. D. Mijic, M. S. Morris, and W.-M. Suen, Phys. Rev. D {\bf 34},
2934 (1986).
\item{16.}
W.-M. Suen and P. R. Anderson, Phys. Rev. D {\bf 35}, 2940 (1987).
\item{17.}
J. B. Hartle, in {\it Gravitation in Astrophysics} ed. B.~Carter
and J.~B.~Hartle (Plenum Press, London, 1987).
\end